\documentclass[a4paper,12pt]{article}

\usepackage{amssymb,amsmath}

\title{Families of classical subgroup separable  superintegrable systems}
\author{%
E.~G.~Kalnins\\Department of Mathematics, University of Waikato,\\Hamilton, New Zealand\\[5mm]%
J.~M.~Kress\\School of Mathematics and Statistics, University if New South Wales,\\ Sydney, Australia\\[5mm]%
W.~Miller,~Jr.\\School of Mathematics, University of Minnesota,\\ Minneapolis, Minnesota, U.~S.~A.%
}
 
\begin{document}

\maketitle

\begin{abstract}
We describe a method for determining a complete set of integrals for a classical Hamiltonian 
that separates in orthogonal subgroup coordinates.  As examples,
we use it  to determine complete sets of integrals, polynomial in the momenta,
for some families of generalized oscillator and Kepler-Coulomb systems, hence
demonstrating their superintegrability. The latter generalizes recent results of Verrier and Evans,  and Rodr\'iguez, Tempesta,  and Winternitz.  Another example is given of a superintegrable
system on a non-conformally flat space.
\end{abstract}

An $n$-dimensional classical superintegrable system is an integrable Hamiltonian 
system that not only possesses $n$ mutually commuting integrals, but in addition,
the Hamiltonian Poisson-commutes with $2n-1$ functions on the phase space that are globally
defined and polynomial in the momenta.  This notion can be extended to define a quantum
superintegrable system as a quantum Hamiltonian which is one of a set
of $n$ independent mutually commuting differential operators that commutes with a set of $2n-1$
independent differential operators of finite order.

Such systems have been studied because of their associated algebras that allow
direct calculation of spectral decompositions and
their connections with separation of variables and special function theory.
There has also been great interest in quantum superintegrable systems because
it has been conjectured that they coincide with quasi-exactly solvable systems \cite{QuasiExact},
 \cite{KMPog1}).

Recently, an infinite
family of exactly solvable quantum mechanical systems was studied by Tremblay, Turbiner
and Winternitz (TTW) \cite{TTW}.  
The proposition that every member of this family of systems
is superintegrable, while not proven, is supported by the fact that their classical counterparts
have been shown to be superintegrable \cite{Ernie}.

In this paper the methods of \cite{KKMPog02} and \cite{Ernie} are used to 
demonstrate that some natural $n$-dimensional generalizations of the TTW class of classical
systems are superintegrable, that is, possess $2n-1$ constants that are polynomial
in the momenta.  All superintegrable systems previously known to us  exist on conformally
flat spaces. However, the underlying metrics of the new systems presented here 
are not necessarily conformally flat, as we show explicitly.

In the first section we discuss a general method for determining
additional constants for a Hamiltonian written in a particular type of
separable coordinates.  This method is then applied to a generalization of the
family of singular oscillator systems given in \cite{TTW} and to a family of
generalized extended Kepler-Coulomb systems. We conclude with an example of a non-conformally flat superintegrable system in four dimensions. Our methods clearly extend to a wide variety of $n$-dimensional systems. Finding the quantum analogs of these systems is a nontrivial problem.

\section{Superintegrability of subgroup separable Hamiltonians}
\label{sec:generaltheory}

Consider the $n$ functions 
\begin{eqnarray}
 L_i &=& p_i^2 + V_i(q_i) + f_i(q_i)L_{i+1}, \qquad i=1,\ldots,n-1, \nonumber \\
 L_n &=& p_n^2 + V_n(q_n),
\label{Ldef}
\end{eqnarray}
on a $2n$-dimensional phase space
with position coordinates $q_i$ and conjugate momenta $p_i$, $i=1,\ldots,n$.
Each $L_i$ is a function of $q_i,\ldots,q_n,p_i,\ldots,p_n$ and
the Hamiltonian $H=L_1$ is clearly separable in
the coordinates $q_i$.
Furthermore, the set $\{L_1,L_2,\ldots,L_n\}$ is in involution, that is
$\{L_i,L_j\}=0$ for all $i,j=1,\ldots,n$, where $\{\ ,\ \}$ is
the usual Poisson bracket,
\[
 \{F,G\} = \sum_{i=1}^n\frac{\partial F}{\partial p_i}\frac{\partial G}{\partial q_i}
 - \frac{\partial F}{\partial q_i}\frac{\partial G}{\partial p_i}.
\]

We can find $n-1$ additional independent functions commuting with $H$ by first
finding, for each $i=1,\ldots,n-1$, a pair of functions $M_i(q_i,L_1,L_2,\ldots,L_n)$ and
$N_i(q_{i+1},L_1,L_2,\ldots,L_n)$ satisfying
\begin{equation}
\label{HMNeqn}
 \{H,M_i\} = \prod_{k=1}^if_k(q_k)
\qquad\mbox{and}
\qquad
\{H,N_i\} = \prod_{k=1}^if_k(q_k).
\end{equation}
Then $\{H,N_i-M_i\}=0$ and hence
\[
 L'_i = N_i-M_i
\]
is in involution with $H$.
These additional functions need not be globally defined nor polynomial
in the momenta,. However, for the examples given in later sections each $L'_i$ gives
rise to an additional constant polynomial in the momenta.

Now,
\[
\{H,M_i\}
= 2p_i\frac{\partial M_i}{\partial q_i}\prod_{k=1}^{i-1}f_k(q_k)
\qquad
\mbox{and}
\qquad
\{H,N_i\}
= 2p_{i+1}\frac{\partial N_i}{\partial q_{i+1}}\prod_{k=1}^{i}f_k(q_k)
\]
and so equations (\ref{HMNeqn}) become
\begin{equation}
\label{MNeqn}
2p_i\frac{\partial M_i}{\partial q_i} = f_i(q_i)
\qquad\mbox{and}\qquad
2p_{i+1}\frac{\partial N_i}{\partial q_{i+1}} = 1
\end{equation}
To solve these equations we note that using (\ref{Ldef}) we can write $p_i$ in
terms of $L_i$ and then treat $L_i$ and $L_{i+1}$ as constants.  Hence we find
\begin{eqnarray}
 M_i &=&  \int\frac{f_i(q_i)dq_i}{2\sqrt{L_i-V_i(q_i)-f_i(q_i)L_{i+1}}} \nonumber \\
 N_i &=& \int\frac{dq_i}{2\sqrt{L_{i+1}-V_{i+1}(q_{i+1})-f_{i+1}(q_{i+1})L_{i+2}}}.
\label{MNintegrals}
\end{eqnarray}
To write these in a consistent form for $i=1,\ldots,n-1$ we take $L_{n+1}=0$.

We also need to check that set $\{L_1,\ldots,L_n,L'_1,\ldots,L'_{n-1}\}$ is indeed functionally
independent.  
First we note that
\begin{align*}
 \{L_i,M_j\} & =  \begin{cases}
                    0 & i>j \\ \displaystyle \prod_{k=i}^jf_k(q_k) & i\leq j
                   \end{cases}
 & \{L_i,N_j\} & =  \begin{cases}
                    0 & i>j+1 \\ 1 & i=j+1 \\ \displaystyle \prod_{k=i}^jf_k(q_k) & i< j+1
                   \end{cases} \\
\end{align*}
and so
\[
 \{L_i,L'_{i-1}\} = 1
\qquad\mbox{for $i>1$}
\qquad\mbox{and}\qquad
 \{L_i,L'_j\} = 0 \qquad \mbox{for $j\neq i-1$.}
\]
If we assume that there exists a function $F$ such that
\begin{equation} \label{Feqn}
 F(L_1,\ldots,L_n,L'_1,\ldots,L'_{n-1}) = 0
\end{equation}
then for each $j=2,\ldots,n$,
\[
 0 = \{L_j,F\} = 
\sum_{k=1}^n \{L_j,L_k\}\frac{\partial F}{\partial L_k} + 
\sum_{k=1}^{n-1} \{L_j,L'_k\}\frac{\partial F}{\partial L'_k}
= \{L_j,L'_{j-1}\}\frac{\partial F}{\partial L'_{j-1}}
= \frac{\partial F}{\partial L'_{j-1}}.
\]
So $F$ can not depend on any of the $L'_i$ and equation (\ref{Feqn}) must
be a functional relationship between the $L_1,\ldots,L_n$ alone which are
are clearly functionally independent.
Hence the set $\{L_1,\ldots,L_n,L'_1,\ldots,L'_{n-1}\}$ must be functionally
independent.

Note that if we were to find an additional function $L'_n$ such that $\{H,L'_n\}=1$, (which we can \cite{KKMPog02})
then the set $\{L_1,\ldots,L_n,L'_1,\ldots,L'_n\}$ would constitute a set of
action-angle variables for the system \cite{Arnold}.

\section{A three-dimensional example}
\label{sec:3dexample}

The details of the calculation for the TTW systems (defined on $E(2,\mathbb C)$) have already been given in \cite{Ernie},
 so here we first consider the Hamiltonian on $E(3,\mathbb C)$
\[
 H = p_x^2 + p_y^2 + p_z^2 + \alpha(x^2+y^2+z^2) + \frac{\beta_1}{z^2} + \frac{\beta_2}{x^2} + \frac{\beta_3}{y^2}
\]
which in polar coordinates is
\[
H = p_r^2 + \frac{p_{\theta_1}^2}{r^2} + \frac{p_{\theta_2}^2}{r^2\sin^2\theta_1} + \alpha r^2 + \frac{\beta_1}{r^2\cos^2\theta_1}
   + \frac{\beta_2}{r^2\sin^2\theta_1\cos^2\theta_2} + \frac{\beta_3}{r^2\sin^2\theta_1\sin^2\theta_2}
\]
and  has the form discussed in section \ref{sec:generaltheory}.
We modify this so that 
\begin{equation}
\label{Hk3d}
H = p_r^2 + \frac{p_{\theta_1}^2}{r^2} + \frac{p_{\theta_2}^2}{r^2\sin^2(k_1\theta_1)} + \alpha r^2
   + \frac{\beta_1}{r^2\cos^2(k_1\theta_1)}.
   + \frac{\beta_2}{r^2\sin^2(k_1\theta_1)\cos^2(k_2\theta_2)}
   + \frac{\beta_3}{r^2\sin^2(k_1\theta_1)\sin^2(k_2\theta_2)}
\end{equation}
with $k_1$ and $k_2$ two positive rational parameters.  Note that when $k_1\neq1$, this is no longer
a natural Hamiltonian on a flat space, however, the Cotton-York tensor of the corresponding
metric vanishes 
identically for all $k_1$ and hence the underlying space is conformally flat.

The Hamilton-Jacobi equation separates due to the second order constants
\[
L_3 = p_{\theta_2}^2 + \frac{\beta_2}{\cos^2(k_2\theta_2)} + \frac{\beta_3}{\sin^2(k_2\theta_2)}
\]
and
\[
L_2 = p_{\theta_1}^2 + \frac{\beta_1}{\cos^2(k_1\theta_1)} + \frac{L_3}{\sin^2(k_1\theta_1)}.
\]
and the Hamiltonian can be written as
\[
H = L_1 = p_r^2 + \alpha r^2 + \frac{L_2}{r^2}.
\]

Now we look for additional constants by finding $M_1(r,H,L_2,L_3)$ and $N_1(\theta_1,H,L_2,L_3)$ such that
\begin{equation} \label{HM1}
\{H,M_1\} = \frac{1}{r^2} \qquad\mbox{and}\qquad \{H,N_1\} = \frac{1}{r^2} 
\end{equation}
and $M_2(\theta_1,H,L_2,L_3)$ and $N_2(\theta_2,H,L_2,L_3)$ satisfying
\begin{equation} \label{HM2}
\{H,M_2\} = \frac{1}{r^2\sin^2(k_1\theta_1)} \qquad\mbox{and}\qquad \{H,N_2\} = \frac{1}{r^2\sin^2(k_1\theta_1)}. 
\end{equation}
Then $L'_1=N_1-M_1$ and $L'_2=N_2-M_2$ will be constants of the motion.  To compute these 
we need to calculate the following 4 integrals,
\begin{eqnarray}
 M_1 &=& \int\frac{dr}{2r^2\sqrt{H-\alpha r^2-\frac{L_2}{r^2}}} \nonumber \\
 N_1 &=& \int\frac{d\theta_1}{2\sqrt{L_2-\frac{\beta_1}{\cos^2(k_1\theta_1)}-\frac{L_3}{\sin^2(k_1\theta_1)}}} \nonumber  \\
 M_2 &=& \int\frac{d\theta_1}{2\sin^2(k_1\theta_1)\sqrt{L_2-\frac{\beta_1}{\cos^2(k_1\theta_1)}-\frac{L_3}{\sin^2(k_1\theta_1)}}}  \nonumber \\
 N_2 &=& \int\frac{d\theta_2}{2\sqrt{L_3-\frac{\beta_2}{\cos^2(k_2\theta_2)}-\frac{\beta_3}{\sin^2(k_2\theta_2)}}},
\label{MNintegrals3d}
\end{eqnarray}
in which $H=L_1$, $L_2$ and $L_3$ are treated as constants.
We find
\[
 M_1 = \frac{{\mathcal B}_1}{4\sqrt{-L_2}},
\quad
 N_1 = \frac{{\mathcal A}_1}{4k_1\sqrt{-L_2}},
\quad
 M_2 = \frac{{\mathcal B}_2}{4k_1\sqrt{-L_3}}
\quad\mbox{and}\quad
 N_2 = \frac{{\mathcal A}_2}{4k_2\sqrt{-L_3}}
\]
where
\begin{align*}
 \sinh{\mathcal B}_1 &= i\ \frac{H-\frac{2L_2}{r^2}}{\sqrt{H^2-4\alpha L_2}}
  & \cosh{\mathcal B}_1 &= - \frac{2\sqrt{L_2}p_r}{r\sqrt{H^2-4\alpha L_2}} \\
\sinh{\mathcal A}_1 &= i\ \frac{L_2\cos(2k_1\theta_1)+L_3-\beta_1}{\sqrt{(\beta_1-L_2-L_3)^2-4L_2L_3}}
  & \cosh{\mathcal A}_1 &= \frac{\sqrt{L_2}\sin(2k_1\theta_1)p_{\theta_1}}{\sqrt{(\beta_1-L_2-L_3)^2-4L_2L_3}} \\
 \sinh{\mathcal B}_2 &=  \frac{2L_3\mbox{cosec}^2(k_1\theta_1)+\beta_1-L_2-L_3}{\sqrt{4\beta_1 L_2-(L_3-L_2-\beta_1)^2}}
  & \cosh{\mathcal B}_2 &= - \frac{2i\sqrt{L_3}\cot(k_1\theta_1)p_{\theta_1}}{\sqrt{4\beta_1 L_2-(L_3-L_2-\beta_1)^2}} \\
 \sinh{\mathcal A}_2 &= i\ \frac{L_3\cos(2k_2\theta_2)+\beta_3-\beta_2}{\sqrt{(\beta_2-\beta_3-L_3)^2-4\beta_3 L_3}}
  & \cosh{\mathcal A}_2 &= \frac{\sqrt{L_3}\sin(2k_2\theta_2)p_{\theta_2}}{\sqrt{(\beta_2-\beta_3-L_3)^2-4\beta_3 L_3}}.
\end{align*}

To ensure that we can find polynomial constants, we have chosen $k_1$ and $k_2$ to be rational and so we can take
\[
 k_1 = \frac{p_1}{q_1} \quad\mbox{and}\quad k_2 = \frac{p_2}{q_2}
\quad\mbox{with}\quad p_1,q_1,p_2,q_2\in\mathbb Z^+
\quad \mbox{and} \quad
\gcd(p_1,q_1)=\gcd(p_2,q_2)=1,
\]
and then
\[
 \sinh\Bigl(4p_1\sqrt{-L_2}(N_1-M_1)\Bigr)
\qquad\mbox{and}\qquad
 \sinh\Bigl(4p_1p_2\sqrt{-L_3}(N_2-M_2)\Bigr)
\]
are constants of the motion having the form of a polynomial in the momenta divided by
a function of $H$, $L_2$, and $L_3$.  This follows from the elementary relations \cite{Ernie}
$$ (\cosh x\pm\sinh x)^n=\cosh nx\pm\sinh nx,\  \cosh (x+y)=\cosh x\cosh y+\sinh x\sinh y,$$
$$   \sinh (x+y)=\cosh x\sinh y+\sinh x\cosh y.$$ 
In particular,
$$ \cosh nx=\sum_{j=0}^{[n/2]} \left( \begin{array}{c}n\\ 2j\end{array}\right) \sinh ^{2j}x\ \cosh^{n-2j}x,$$
$$  \sinh nx=\sinh x\sum_{j=1}^{[(n=1)/2]} \left( \begin{array}{c}n\\ 2j-1\end{array}\right) \sinh ^{2j-2}x\ \cosh^{n-2j-1}x.$$
From these we can find
 $L''_1$ and $L''_2$ such that $\{H,L_2,L_3,L''_1,L''_2\}$ is a set of
functionally independent constants of the motion polynomial in the momenta.

Note these polynomial constants are not necessarily of minimal degree.  Indeed, if we
set $k_1=k_2=1$ and recover the usual Smorodinski-Winternitz potential, we find additional
constants $L''_1$ and $L''_2$ that are cubic in the momenta, whereas it is well known that
there exist additional quadratic constants.

\section{Extended Kepler-Coulomb system}
\label{sec:KeplerCoulomb}

The same procedure works with the harmonic oscillator term $\alpha r^2$ in (\ref{Hk3d}) replaced
by a Kepler-Coulomb term $\alpha/r$, that is,
\begin{equation}
\label{Hk3dKC}
H = p_r^2 + \frac{p_{\theta_1}^2}{r^2} + \frac{p_{\theta_2}^2}{r^2\sin^2(k_1\theta_1)} + \frac\alpha r
   + \frac{\beta_1}{r^2\cos^2(k_1\theta_1)}.
   + \frac{\beta_2}{r^2\sin^2(k_1\theta_1)\cos^2(k_2\theta_2)}
   + \frac{\beta_3}{r^2\sin^2(k_1\theta_1)\sin^2(k_2\theta_2)}
\end{equation}
This system is superintegrable for rational $k_1$ and $k_2$.  
All calculations are the same as for the singular oscillator except
\[
M_1 = \int\frac{dr}{2r^2\sqrt{H-\frac{\alpha}r-\frac{L_2}{r^2}}} \\
\]
and
\[
 M_1 = \frac1{2\sqrt{-L_2}} \sinh^{-1}\left(i \frac{\alpha+\frac{2L_2}{r}}{\sqrt{\alpha^2+4HL_2}}\right)
\]
or
\[
 M_1 = \frac{{\mathcal B}_1}{2\sqrt{-L_2}}
\qquad\mbox{where}\qquad
\sinh{\mathcal B}_1 = i\frac{\alpha+\frac{2L_2}{r}}{\sqrt{\alpha^2+4HL_2}}
\qquad\mbox{and}\qquad
\cosh{\mathcal B}_1 = \frac{2\sqrt{L_2}p_r}{\sqrt{\alpha^2+4HL_2}}
\]
Verrier and Evans \cite{VE2008}, and Rodr\'iguez, Tempesta,  and Winternitz \cite{RTW}, separately, considered the $k_1=k_2=1$ case of this family of systems and
found it to be superintegrable with 4 second order and one fourth order constant.  The method
presented here leads to, in addition to the 3 second order constants $H$, $L_2$ and $L_3$, a
third and a fourth order constant. 


\section{A non-flat higher-dimensional example}
\label{sec:}

The three-dimensional examples above can readily be extended to $n$ dimensions to give families of
$n$-dimensional superintegrable Hamiltonians, each with $2n-1$ functionally independent
polynomial constants of the momenta.  In 4-dimensions, by calculating the Weyl tensor, it can be seen
that the corresponding metric is
conformally flat if and only if $k_1=k_2$.  Hence we can generate examples of superintegrable
systems on non-conformally flat spaces.


As an example consider
the natural generalization of the example above to 4 dimensions with $k_1=2$ and $k_2=k_3=1$.
That is,
\begin{eqnarray*}
L_4 &=& p_{\theta_3}^2 + \frac{\beta_3}{\cos^2(\theta_3)} + \frac{\beta_4}{\sin^2(\theta_3)} \\
L_3 &=& p_{\theta_2}^2 + \frac{\beta_2}{\cos^2(\theta_2)} + \frac{L_4}{\sin^2(\theta_2)} \\
L_2 &=& p_{\theta_1}^2 + \frac{\beta_1}{\cos^2(2\theta_1)} + \frac{L_3}{\sin^2(2\theta_1)} \\
H = L_1 &=& p_r^2 + \alpha r^2 + \frac{L_2}{r^2}.
\end{eqnarray*}

As in the previous example, we calculate $M_1(r,H,L_2,L_3,L_4)$, $N_1(\theta_1,H,L_2,L_3,L_4)$, 
$M_2(\theta_1,H,L_2,L_3,L_4)$, $N_2(\theta_2,H,L_2,L_3,L_4)$, 
$M_3(\theta_2,H,L_2,L_3,L_4)$ and $N_3(\theta_3,H,L_2,L_3,L_4)$ using (\ref{MNintegrals}) to
give very similar expressions.  We can form polynomial constants from $L'_i=N_i-M_i$ for 
$i=1,2,3$.

\begin{eqnarray*}
\sinh(8\sqrt{-L_2}(N_1-M_1))
 &=& \sinh({\mathcal A}_1 - 2{\mathcal B}_1) \\
 &=& -2\cosh{\mathcal A}_1\sinh{\mathcal B}_1\cosh{\mathcal B}_1 + 2\sinh{\mathcal A}_1\cosh^2{\mathcal B}_1 - \sinh{\mathcal A}_1\\
 &=& \frac{\displaystyle4iL_2\left( \left(H-\frac{2L_2}{r^2}\right)\frac{\sin(4\theta_1)}{r}p_{\theta_1}p_r + \frac{2(L_2\cos(4\theta_1)+L_3-\beta_1)}{r^2}\ p_r^2\right)}{(H^2-4\alpha L_2)\sqrt{(\beta_1-L_2-L_3)^2-4L_2L_3}} \\
 \\ & & \quad {} - i\ \frac{(H^2-4\alpha L_2)(L_2\cos(4\theta_1)+L_3-\beta_1)}{(H^2-4\alpha L_2)\sqrt{(\beta_1-L_2-L_3)^2-4L_2L_3}}. \\
\end{eqnarray*}
The denominator of this expression is a constant of the motion and hence so is the numerator which is
clearly a polynomial in the momenta.
We also can find a lower degree constant $L''_1$ by noting that
\[
 \sinh(8\sqrt{-L_2}(N_1-M_1))=\frac{4iL_2L''_1 - iH^2(L_3-\beta_1)}{(H^2-4\alpha L_2)\sqrt{(\beta_1-L_2-L_3)^2-4L_2L_3}}
\]
where
\[
 L''_1 =  \left(H-\frac{2L_2}{r^2}\right)\frac{\sin(2\theta_1)}{r}p_{\theta_1}p_r + \frac{2(L_2\cos(2\theta_1)+L_3-\beta_1)}{r^2}\ p_r^2 - \frac14(H^2-\alpha L_2)\cos(4\theta_1)
\]
is a constant quartic in he momenta.  

Similarly,
\begin{eqnarray*}
\lefteqn{\sinh(8\sqrt{-L_3}(N_2-M_2))} \\
 &=& \sinh(2{\mathcal A}_2 - {\mathcal B}_2) \\
 &=& -2\cosh^2{\mathcal A}_2\sinh{\mathcal B}_2 + 2\sinh{\mathcal A}_2\cosh{\mathcal A}_2\cosh{\mathcal B}_2 + \sinh{\mathcal B}_2\\
&=& L_3\ \frac{
    2(L_3\cos(2\theta_2)+L_4-\beta_2)\cot(2\theta_1)\sin(2\theta_2)p_{\theta_1}p_{\theta_2}
    - \sin^2(2\theta_2)(2L_3\mbox{cosec}^2(2\theta_1)+\beta_1-L_2-L_3)p_{\theta_2}^2
       }{((\beta_2-L_3-L_4)^2-4L_3L_4)\sqrt{4\beta_1 L_2-(L_3-L_2-\beta_1)^2}} \\
& & \quad {}
       + \frac{((\beta_2-L_3-L_4)^2-4L_3L_4)(2L_3\mbox{cosec}^2(2\theta_1)+\beta_1-L_2-L_3)}{((\beta_2-L_3-L_4)^2-4L_3L_4)\sqrt{4\beta_1 L_2-(L_3-L_2-\beta_1)^2}}. \\
\end{eqnarray*}
Hence
\[
 L''_2 =  
    2(L_3\cos(2\theta_2)+L_4-\beta_2)\cot(2\theta_1)\sin(2\theta_2)p_{\theta_1}p_{\theta_2}
 + ((\beta_2-L_3-L_4)^2-4L_3L_4) \mbox{cosec}^2(2\theta_1)
\]
\[
   - \sin^2(2\theta_2)(2L_3\mbox{cosec}^2(2\theta_1)+\beta_1-L_2-L_3)p_{\theta_2}^2
\]
is an additional constant that is quartic in the momenta, and
\begin{eqnarray*}
\lefteqn{\sinh(4\sqrt{-L_4}(N_3-M_3))} \\
 &=& \sinh({\mathcal A}_3 - {\mathcal B}_3) \\
 &=& \sinh{\mathcal A}_3\cosh{\mathcal B}_3 - \cosh{\mathcal A}_3\sinh{\mathcal B}_3 \\
 &=& \frac{\sqrt{L_4}\Bigl(
       2 (L_4\cos(2\theta_3)+\beta_4-\beta_3)\cot(\theta_2)p_{\theta_2}
       -(2L_4\mbox{cosec}^2(\theta_2)+\beta_2-L_3-L_4)\sin(2\theta_3)p_{\theta_3}
          \Bigr)}{\sqrt{4\beta_2L_3-(L_4-L_3-\beta_2)^2}\sqrt{(\beta_3-\beta_4-L_4)^2-4\beta_4L_4}}.
\end{eqnarray*}
Hence
\[
 L''_3 =
        2 (L_4\cos(2\theta_3)+\beta_4-\beta_3)\cot(\theta_2)p_{\theta_2}
      -(2L_4\mbox{cosec}^2(\theta_2)+\beta_2-L_3-L_4)\sin(2\theta_3)p_{\theta_3}
\]
is an additional constant that is cubic in the momenta.

So we have demonstrated the superintegrability of one member of this family of superintegrable 
systems by giving explicit expressions for $2n-1$ functionally independent polynomial 
constants.  In this case, the underlying space is curved and not conformally flat as
has been the case for previously known superintegrable systems.

It is clear from this example, that similar results will be obtained for both this family
of generalized oscillators and the family of generalized Kepler-Coulomb
systems for any rational
choices of the $k_1, k_2,\ldots,k_{n-1}$ in $n$ dimensions.

\label{thelastpage}

\end{document}